\documentclass[
reprint,
superscriptaddress,
amsmath,
amssymb,
aps
]{revtex4-1}

\usepackage{orcidlink}
\usepackage{academicons}
\usepackage{amsmath}
\usepackage{hyperref}
\usepackage{mathtools}
\usepackage{xcolor}
\usepackage{tensor}
\usepackage{float}
\usepackage{comment}
\usepackage{caption}
\usepackage{subcaption}
\usepackage{tikz}
\usepackage{tzplot}

\begin{document}

\title{Detecting cosmological scalar fields using orbital networks of quantum sensors}
\author{Yu Li \orcidlink{0009-0009-1565-6356}}
\affiliation{Waterloo Centre for Astrophysics, University of Waterloo, Waterloo, ON, N2L 3G1, Canada}
\affiliation{Perimeter Institute for Theoretical Physics, Waterloo, ON, N2L 2Y5, Canada}
\affiliation{Department of Applied Mathematics, University of Waterloo, Waterloo, ON, N2L 3G1, Canada}
\author{Ruolin Liu\orcidlink{0009-0003-7385-3071}}
\affiliation{Waterloo Centre for Astrophysics, University of Waterloo, Waterloo, ON, N2L 3G1, Canada}
\affiliation{Department of Physics and Astronomy, University of Waterloo, Waterloo, ON, N2L 3G1, Canada}
\affiliation{Perimeter Institute for Theoretical Physics, Waterloo, ON, N2L 2Y5, Canada}
\author{Conner Dailey\orcidlink{0000-0003-2488-3461}}
\affiliation{Waterloo Centre for Astrophysics, University of Waterloo, Waterloo, ON, N2L 3G1, Canada}
\affiliation{Department of Physics and Astronomy, University of Waterloo, Waterloo, ON, N2L 3G1, Canada}
\affiliation{Perimeter Institute for Theoretical Physics, Waterloo, ON, N2L 2Y5, Canada}
\author{Niayesh Afshordi \orcidlink{0000-0002-9940-7040}}
\affiliation{Waterloo Centre for Astrophysics, University of Waterloo, Waterloo, ON, N2L 3G1, Canada}
\affiliation{Department of Physics and Astronomy, University of Waterloo, Waterloo, ON, N2L 3G1, Canada}
\affiliation{Perimeter Institute for Theoretical Physics, Waterloo, ON, N2L 2Y5, Canada}
\date{\today}

\begin{abstract}
In this paper, we propose a method to detect the interaction of a hypothetical coherently evolving cosmological scalar field with an orbital network of quantum sensors, focusing on the GPS satellite network as a test example. Cosmological scenarios, such as a scalar-tensor theory for dark energy or the axi-Higgs model, suggest that such a field may exist. As this field would be (approximately) at rest in the CMB frame, it would exhibit a dipole as a result of the movement of our terrestrial observers relative to the CMB. While the current sensitivity of the GPS network is insufficient to detect a cosmological dipole, future networks of quantum sensors on heliocentric orbits, using state-of-the-art atomic clocks, can reach and exceed this requirement.    
\end{abstract}

\maketitle

\section{Introduction}
One of the most successful predictions of the cosmological Big Bang theory has been the existence of a relic gas of radiation, first discovered as cosmic microwave background (CMB) \cite{penzias1}. This profound discovery and the subsequent measurement of its dipole associated with our motion relative to the CMB frame \cite{1969Natur.222..971C} have been our first tangible exposure to a cosmological reference frame, on scales that are at least 20 orders of magnitude larger than our planet. But is it possible to have other local probes of this cosmic frame on Earth? For example, the cosmic neutrino background, also predicted by Big Bang theory, remains out of experimental reach for now because of the weak interaction of relic neutrinos. In this Letter, we instead explore the potential for detecting a coherent cosmic scalar field, based on an entirely different technology.     

In recent work \cite{Dailey_2020,Roberts_2017,Roberts_2018,Afach_2021} there has been an interest in signal candidates observable by a network of quantum sensors, such as the network of atomic clocks that makes up the Global Positioning System (GPS) or the Global Network of Optical Magnetometers for Exotic Physics (GNOME). The signal candidates that have been proposed tend to be transient signals that pass through the network, such as the galactic domain walls that are coupled to GPS \cite{Roberts_2017} or GNOME \cite{Afach_2021}. There have also been proposals to search for other exotic physics using these networks, such as extragalactic exotic low-mass fields (ELFs) \cite{Dailey_2020}, which may be emitted by high energy astrophysical events such as Binary Black Hole mergers (BBHs) and Binary Neutron Star mergers (BNSs), with comparable wavefroms to those detected by gravitational wave observatories.  

In contrast, here we consider a candidate that is not transient but rather sourced by a coherent cosmological scalar field such as those suggested by axi-Higgs cosmology \cite{axi-Higgs}, or coupling to a quintessence dark energy fluid (e.g., \cite{zlatev1999quintessence,chiba2002quintessence}). Such a field would evolve slowly in the frame of the CMB, and thus, as seen by us on Earth, would exhibit a dipole due to the velocity of the Earth through the field. We propose to detect this coherent dipole by assuming various couplings detectable by atomic clocks or magnetometers.

On a separate front, the nature of the CMB dipole has been a topic of considerable fascination among cosmologists over the past few decades. For example, it is not clear that the gravity of the observed large-scale structure can entirely explain the velocity of the local group (which is the largest contributor to the CMB dipole), in the context of the standard cosmological model (e.g., \cite{lavaux2010cosmic}). Moreover, the kinematic dipole inferred from the distribution of distant galaxies and quasars is arguably in conflict with the CMB dipole \cite{secrest2021test,secrest2022challenge} (but see \cite{dalang2022kinematic}). Due to the special nature of constant acceleration in the general equivalence principle, it is not easy to explain such an anomaly without introducing new physics (e.g., \cite{erickcek2008superhorizon,domenech2022galaxy,krishnan2023dipole}). Therefore, an independent measurement of the cosmological dipole may shed light on the nature of these anomalies.   

As the average dipole signal from this scalar field over a year would be approximately constant in direction and time, a long integration period from a network of quantum sensors, such as the 25 years of GPS data publicly available, can act to make sensitive measurements. We shall first start by motivating the theoretical scenarios that we shall consider for our exploration.

\section{cosmological scalar fields}

Although the simplest type of quantum field theories involve scalar fields, the only fundamental scalar field observed in nature so far is the Higgs field. Nonetheless, light scalar fields are often invoked as simple possible models for inflation, dark matter, and/or dark energy in cosmology. Furthermore, a variety of light pseudo-scalar fields (of different masses), i.e., axions, may be expected from theories such as string theory (e.g., \cite{Arvanitaki_2010}). Such light scalars are expected to evolve coherently in the cosmological rest frame, if they are stable and their mass is comparable to the Hubble rate ($\sim 10^{-33}$ eV today)\footnote{In contrast, heavier fields cluster and may act as dark matter}. For example, if such a light field is responsible for the recent onset of the acceleration of cosmic expansion, it may play the role of a dynamical dark energy. Current cosmological observations place an upper limit on the rate of change of this field $\dot{\phi}$, compared to the energy density $\rho_{DE}$ and pressure $p_{DE}$ of dark energy (e.g., \cite{Afshordi:2015smm}):
\begin{align}
    \frac{1}{2}\dot{\phi}^2 < \rho_{\rm DE}+p_{\rm DE}  \lesssim (1.3 ~{\rm meV})^4\,. \label{eq2}
\end{align}
While such a field may only couple gravitationally to standard model, it is also possible that it directly couples to local clocks and rulers. 

A concrete example is the axi-Higgs model that may ease certain tensions in the standard cosmological model \cite{axi-Higgs}, i.e., the Hubble tension (e.g., \cite{Di_Valentino_2021}) and the Lithium problem (e.g., \cite{doi:10.1146/annurev-nucl-102010-130445}), by shifting the vacuum expectation value (VEV) of the Higgs field by $\sim 1\%$ in the early universe. 

In this model, the Higgs VEV, $v$, is modulated by an axion field $\phi$  \cite{axi-Higgs}:
\begin{align}
    \frac{\delta v}{v}=C\frac{\phi^2}{2M^2_{\rm Pl}}\label{Eq2}\,,
\end{align}
where $M_{\rm Pl} \simeq 2.4 \times 10^{18} $ GeV is the reduced Planck mass \cite{axihubble} and $C$ is the coupling constant of the Higgs with axion \cite{Hoang2023}. 
In the single-axion model, the axion's evolution is governed by a damped harmonic oscillator:
\begin{align}
    \ddot{\phi}+3H(t)\dot{\phi}+\frac{\partial V}{\partial \phi}=0\,,
\end{align}
where $V$ is the total scalar potential, this evolution equation has a general solution:
\begin{align}
    \phi(t) = A(t)\phi_{0}\cos(m_{\phi}t)\,, \label{Eq7}
\end{align}
where $A(t)$ is a dimensionless amplitude that exponentially decreases with reciprocal 
Hubble constant $H(t)^{-1}$, $\phi_0$ is the initial amplitude of the field, and $m_\phi$ is the mass of the field in natural units.

There are a variety of couplings that can render the field $\phi$ (such as axi-Higgs) detectable by networks of atomic clocks. Common ones proposed include
\begin{align}
    \mathcal{L}_{\mathrm{int}}=\phi^2\left(\Gamma_f m_fc^2\bar\psi_f\psi_f+\frac{\Gamma_\alpha}{4}F^{\mu\nu}F_{\mu\nu}\right)\,.
\end{align}
Here, the subscript $f$ sums over all fermion fields $\psi_f$ and describes a quadratic coupling of $\phi$ to the mass terms of the fermions, as well as a quadratic coupling to the electromagnetic term, with coupling constants $\Gamma_X$. As discussed in \cite{Roberts_2017, Roberts_2018}, these couplings lead to an apparent modulation of fundamental constants, such as the fermion masses $m_f$ and the fine structure constant $\alpha$:
\begin{align}
    \alpha^{\rm eff} &\approx (1+\Gamma_\alpha\phi^2)\alpha\,,\\
    m_f^{\rm eff} &= (1+\Gamma_f\phi^2)m_f\,.
\end{align}
Since atomic transition frequencies are highly sensitive to the masses of the proton $m_p$ and electron $m_e$, as well as the strength of the electromagnetic field governed by $\alpha$, this will lead to an apparent shift in an atomic clock frequency.

Magnetometers couple instead to the spins of fermions \cite{Afach_2021}:
\begin{align}
    \mathcal{L}_{\mathrm{int}}=\frac{(\hbar c)^{3/2}}{f_l}\bar\psi_f\gamma^\mu\gamma^5\psi_f\partial_\mu\phi\,,
\end{align}
where the strength of the coupling is modulated by $f_l$. This term describes a coupling of the gradient of $\phi$ to the fermionic spin terms in the standard model Lagrangian.  

\section{Dipole Modulation of Frequency}\label{modulation to frequency}

Although we consider many possible couplings above, let us first consider only couplings to atomic clocks as an example. These couplings will cause the transition frequencies of atomic clock sensors of type $a$ to have a uniform time dependence in the CMB frame:
\begin{align}
\frac{\delta \nu_a({\bf x}, t)}{\nu} = \Gamma^{\rm eff}_a \delta \phi^2(t_{\rm CMB}) = g_a \delta \phi(t_{\rm CMB})\,,     \label{eq1}
\end{align}
where $g_a \equiv 2 \Gamma^{\rm eff}_a \phi(t_{\rm CMB})$ quantifies the effective coupling of the sensor to the small variation of the cosmological field. 

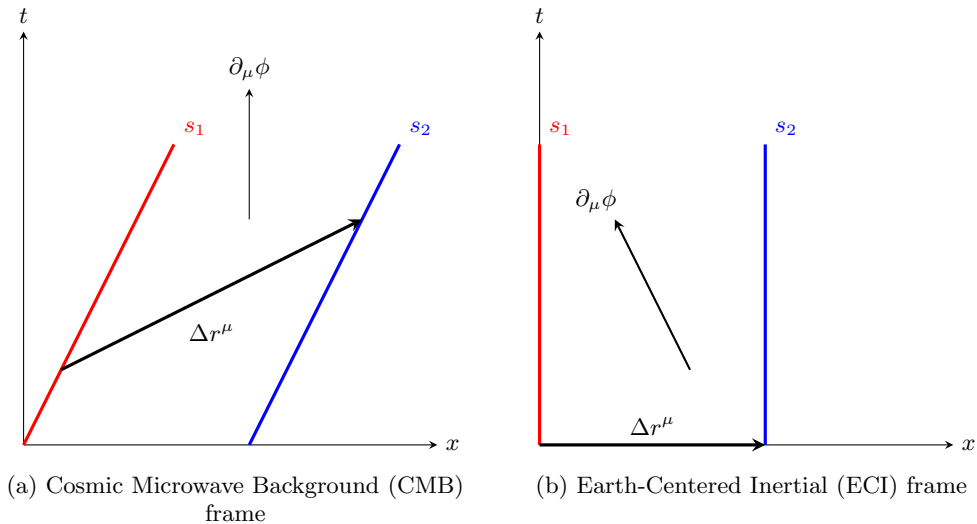
\begin{figure*}
\centering
\begin{subfloat}[Cosmic Microwave Background (CMB) frame]{
\centering
\begin{tikzpicture}
\tzaxes(0,0)(5.5,5.5){$x$}{$t$}
\tzfn[red,very thick]{2*\x}[0:2]{$s_1$}[ar]
\tzfn[blue,very thick]{2*\x-6}[3:5]{$s_2$}[ar]

\node (A) at (3, 3) {};
\node (B) at (3, 5) {$\partial_{\mu}\phi$};
\node (C) at (2.5, 1.5) {$\Delta r^{\mu}$};

\draw[->, to path={-| (\tikztotarget)}]
  (A) edge (B) ;

\draw[->, very thick] (0.5,1) -- (4.5,3);

\end{tikzpicture}
}
\end{subfloat}
\hspace{10pt}       
\begin{subfloat}[Earth-Centered Inertial (ECI) frame]{
\centering
\begin{tikzpicture}
\tzaxes(0,0)(5.5,5.5){$x$}{$t$}
\tzfn'[red,very thick]{0}[0:4]{$s_1$}[ar]
\tzfn'[blue,very thick]{3}[0:4]{$s_2$}[ar]

\node (B) at (0.75, 3.25) {$\partial_{\mu}\phi$};
\node (C) at (1.5, 0.25) {$\Delta r^{\mu}$};

\draw[->, very thick] (0,0) -- (3,0);
\draw[->, thick] (2,1) -- (1,3);

\end{tikzpicture}
}       
\end{subfloat}
\caption{ Spacetime diagrams of two satellites $s_1$ and $s_2$, their separation vector $\Delta r^\mu$ and the gradient of the cosmic scalar that modulates their clocks $\partial_\mu \phi$. The left (right) panel shows the vectors in CMB (ECI) frame. }
\label{fig3}

\end{figure*}

Comparing different transitions of terrestrial atomic clocks already puts limits on $|(g_a-g_{a'}) \dot{\phi}|$ \cite{godun2014frequency}:
\begin{align}
| (g_a-g_{a'}) \dot{\phi}| \lesssim  10^{-16}~ {\rm yr}^{-1},    \label{eq13}
\end{align}
where $g_a$ and $g_{a'}$ are for couplings to transitions $a$ and $a'$ to the $\phi$ field (namely 467 nm and 436 nm transitions of $^{171}$Yb$^+$ in  \cite{godun2014frequency}). 

However, this bound is not applicable if the coupling to $\phi(t)$ does not change the ratio of different atomic transition frequencies at the same location (i.e., $g_a=g_{a'}$).  As such, it makes sense to explore a purely spatial modulation of these transitions. %\todo{This entire paragraph seems out of place. Can we move it somewhere else?}

The coupling between the transition frequencies of the sensors and the cosmological scalar field yields a 
dipole in the satellite data. To see this, consider the Lorentz transformation between \(t_{\rm CMB}\)
and Earth time \(t_{\rm E}\):
\begin{equation}
    t_{\rm CMB} = \gamma \left(t_{\rm E} + \frac{ v^i_{\rm CMB} r_i}{c^2}\right)\,, \label{eq4}
\end{equation}
where $ v^i_{\rm CMB}$ is the velocity of the Earth relative to the cosmic frame and $ r_i$ is the position of the satellite in the Earth-Centered Inertial (ECI) frame. Here, we have ignored the relative motion of the satellites to Earth, since $v_{\rm CMB}$ is dominant (\(v_{\rm CMB} \sim 400~ \text{km/s}\) and \(v_{\text{sat}} \sim 3~ \text{km/s}\)). Since \(v_{\rm CMB} \ll c\), we have
\begin{equation}
    \phi(t_{\rm CMB} ) = \phi\left(t_{\rm E} + \frac{ v^i_{\rm CMB} r_i}{c^2}\right)
    \approx \phi(t_{\rm E}) + \dot{\phi}(t_{\rm E}) \frac{ v^i_{\rm CMB} r_i}{c^2}\,.
    \label{eq5}
\end{equation}
When substituted into Eq.~(\ref{eq1}), we obtain an expression for the modulation of the GPS atomic clock frequencies:
\begin{align}
    \frac{\delta \nu_a({ r^i}, t)}{\nu} =  \frac{\delta \nu_a({ r}^i_{\rm E}, t_{\rm E})}{\nu}  + g_a\dot{\phi}(t_{\rm E}) \frac{{ v^i_{\rm CMB}} { r_i} }{c^2} \nonumber\\ \simeq  \frac{\delta \nu_a({ r}^i_{\rm E}, t_{\rm E})}{\nu}  + g_a\partial_\mu\phi(t_{\rm E}) \Delta r^\mu \,.\label{eq6}
\end{align}
In the last line, we wrote the relative difference clock frequencies in an explicit Lorentz-invariant form, where $\Delta r^\mu$ is the space-like 4-vector, which connects one satellite's worldline to another, and is normal to their 4-velocities. Figure (\ref{fig3}) shows the spacetime diagram of these vectors in both CMB and ECI frames. 

\begin{figure*}
    \centering
    \includegraphics[width=1.0\textwidth]{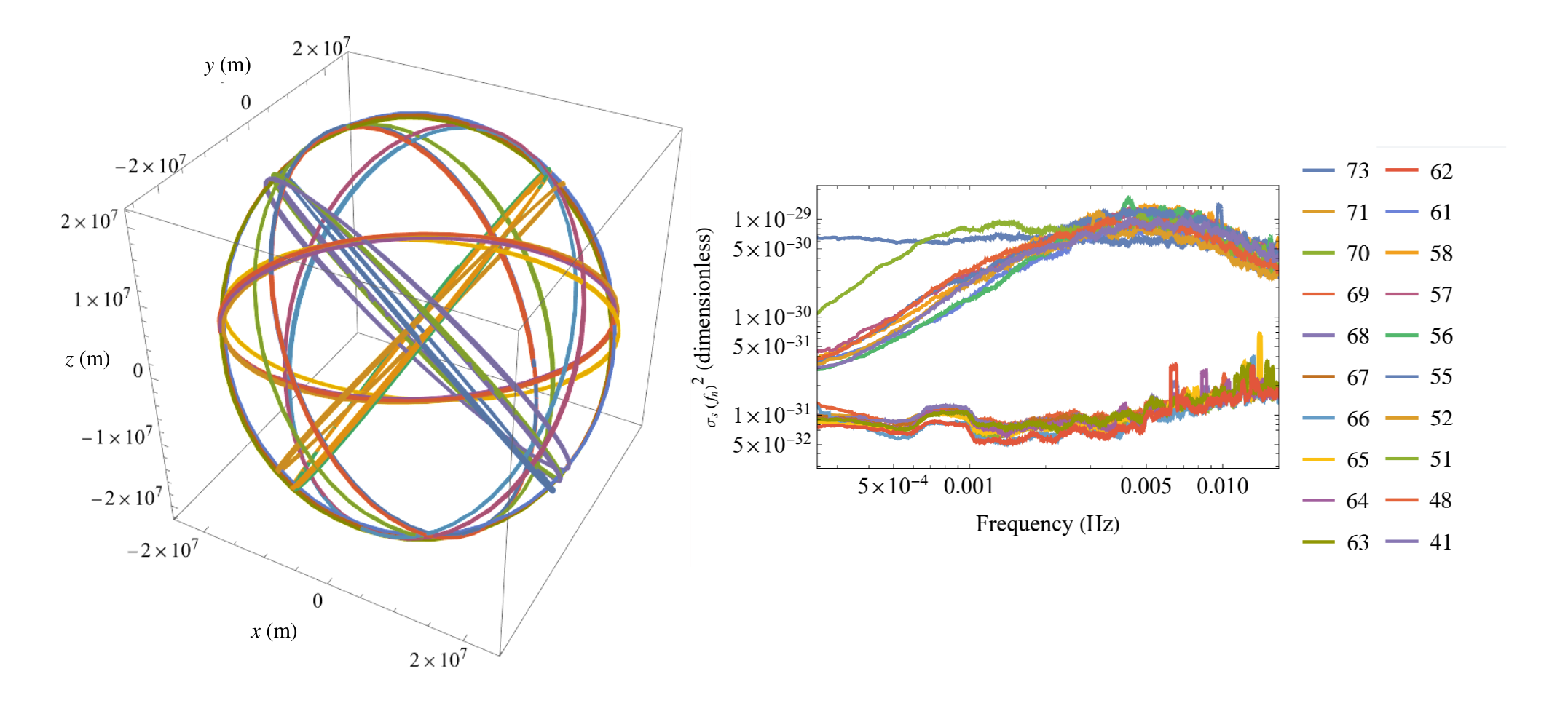}
    \caption{Left: The orbits of the GPS satellites in 3D space in the ECI reference frame. Right: Noise power spectrum of each satellite, calculated based on the GPS public timing bias data during the time period of July 20th to 29th, 2023. The noise power spectrum has been smoothed by taking the moving average with \(\Delta f = 0.185\, \mathrm{mHz}\). The plot legends label the satellite vehicle number of each satellite.}
    \label{fig1}
\end{figure*}
\section{GPS Data Analysis}

Let us briefly discuss the type of information provided by the atomic clocks on top of GPS constellations. 

Available publicly from the GPS division of the Jet Propulsion Laboratory (JPL), the daily clock solutions are a least squares optimization solution that yields the Cartseian positions and timing bias of each GPS satellite,  at each measurement sample interval. The solutions also subtract out several known physical phenomena, such as time dilation and ionospheric effects. The timing bias is the difference in clock phase between a specific satellite's clock and a common reference clock, as yielded by the solution procedure above (see \cite{BLEWITT2015307,Roberts_2018} for a review). The basic idea is that if we have clock phase measurements taken (via microwave communication) by {\it more} than $4$ terrestrial stations for each satellite, we have enough information to solve for both the timing bias and the positions of the satellites. Therefore, we can search for a position-dependent modulation of timing bias, as a result of coupling to a cosmological scalar field, even for a single type of clock.

Let us denote the recorded timing bias as $d^s_n$ for the satellite $s$ and time sample index $n$. This data can be expressed as an integral over the time sample interval
\begin{equation}
    d^s_n = c\int^{t_n}  \frac{\delta \nu^s(t)}{\nu}~dt\,. \label{eq14}
\end{equation}
Note that $d^s_n$ has the dimension of length, which is the convention used in the original timing bias data from NASA\cite{NASAdata4}.
Eq.~(\ref{eq14}) allows the transition frequency change to be approximated by a first-order finite differencing scheme:
\begin{equation}
    \frac{\delta \nu^s(t_n)}{\nu}  \approx \frac{ d^s_n -d^s_{n-1}}{c (t_n-t_{n-1})}\,. \label{dnu}
\end{equation}
To begin our analysis, we define $\chi^2$ for the comparison of our dipole model with data, assuming uncorrelated noise in the Fourier domain:
\begin{align}
    \chi^2 = \sum_{s, n} \frac{\left|q^i \tilde{r}^{s}_i(f_n)  -  \frac{\delta \tilde{\nu}^s(f_n)}{\nu} \right|^2}{\sigma^2_s(f_n)}\,,  \label{eq16}
\end{align}
where 
\begin{align}
    \sigma^2_s(f_n) \equiv \left\langle \left| \frac{\delta \tilde{\nu}^s (f)}{\nu} \right|^2\right\rangle_{f_n \pm \Delta f}\,,~~ q^i \equiv g_a\dot{\phi}\frac{v^i_{{\rm CMB}}}{c^2}\,. 
    \label{variance}
\end{align}
Here, $\delta \tilde{\nu}^s(f)$ is the Fourier transform of $\delta \nu^s (t)$, and $\sigma^2_s(f_n)$ is its band-limited variance within the frequency range ${(f_n-\Delta f,f_n+\Delta f)}$. Similarly, $\tilde{r}^{s}_i(f)$ is the Fourier transform of the satellite position in the ECI frame $ r^s_i(t)$. The ECI frame is used in the satellite position data from NASA\cite{NASAdata4}. This choice of reference frame seems arbitrary, but ultimately the frame choice should not matter since the monopole of the orbital trajectory in ECI is zero. In other words, the difference between the position of the target and the reference clock modulates the dipole effect \footnote{Since the radius of Earth is smaller than the orbits of satellites by a factor of 4, the dipolar modulation of Earth-based reference clock is a subdominant contribution to the model and is ignored in this preliminary analysis.}. 

Finding the minimum value of $\chi^2$, the mean and the covariance of $q_i$ can be determined:
\begin{equation}
    \langle q^i \rangle = (A^{-1})^{i j} B_j\,,\quad \langle \Delta q^i \Delta q^j \rangle  = (A^{-1})^{i j}, \label{eq18}
\end{equation}
where the Fisher matrix is given by
\begin{equation}
    A_{ij}=\sum_{s,n}\frac{\tilde{r}_i^{ s}(f_n)\,\tilde{r}_j^{ s *}(f_n)}{\sigma^2_s(f_n)}\,,
\end{equation}
and
\begin{equation}
    B_{i}=\mathrm{Re}\sum_{s,n}\frac{\tilde{r}_i^{ s}(f_n)}{\sigma^2_s(f_n)} \frac{\delta \tilde{\nu}^{s*}(f_n)}{\nu} \,. 
\end{equation}
Note that, according to the definition of the $q^i$ variables in Eq.~(\ref{variance}), the three-vector $q^i$ is predicted to be in the same direction as $v^i_{\text{CMB}}$. However, treating $q^i$ as a free vector (with the best-fit value and its covariance given in Eq.~(\ref{eq18}) provides an additional sanity check for the consistency of the direction of measured $q^i$ with $v^i_{\text{CMB}}$, which could otherwise be impacted by unknown systematics. However, at the expense of losing sight of such potential systematics, we may achieve a smaller error by assuming $q^i \propto v^i_{\text{CMB}}$:
\begin{equation}
    {g_a \dot \phi}=\frac{v^i_{{\rm CMB}} B_i c^2}{A_{i j}v^i_{{\rm CMB}} v^j_{{\rm CMB}} } \pm \frac{c^2}{\sqrt{A_{i j}v^i_{{\rm CMB}} v^j_{{\rm CMB}}} }\,.
\end{equation}
Ultimately, a reliable detection requires both high statistical significance, and passing all the sanity checks, such as pointing in the right direction as the cosmological dipole.
%\textcolor{blue}{Alternatively, if the error of the best-fit value of $q^i$ is large so that a real detection cannot be ascertained, the error term in the above equation instead imposes bounds on the value of $g_a\dot{\phi}$. 

\section{Results}

We have performed the above linear regression technique in Fourier space with the publicly available GPS timing bias (sampled every \(30 \text{ s}\)) and GPS position data \cite{NASAdata4}.
Note that we have not used all satellites in the GPS constellation for data analysis, as the publicly available timing bias data in \cite{NASAdata4} have missing data points during our time range of interest, so we have chosen satellites with complete data for this analysis. Figure~\ref{fig1} (left) shows the orbits of these GPS satellites in the ECI frame.

Another remark is that we fit the data in frequency space because the noise in real time is correlated, as the noise power spectrum $\sigma^2_s(f_n)$ [Figure~\ref{fig1} (right)], is not constant. However, as long as the noise is statistically invariant under time translations, it will remain uncorrelated in frequency, which justifies the use of Eq. \ref{eq16}  for $\chi^2$ in the Gaussian approximation. 

As a concrete example, we can analyze the GPS data for the half-day data extracted for July 29th, 2023:
\begin{align}
    &\overline{q}^x = ( -2.23\pm 1.76)\times 10^{-25}\, \mathrm{m}^{-1}\nonumber\,, \\ 
    &\overline{q}^y = (+ 0.21 \pm 1.41) \times 10^{-25}\, \mathrm{m}^{-1} \nonumber\,, \\  
    &\overline{q}^z = (+ 3.34 \pm 1.64 ) \times 10^{-25}\, \mathrm{m}^{-1}\,.
\end{align}
Extending this to the full day yields smaller errors:
\begin{align}
    &\overline{q}^x = (-1.02 \pm 1.30)\times 10^{-25}\, \mathrm{m}^{-1}\nonumber\,, \\ 
    &\overline{q}^y = (+1.56 \pm 1.48) \times 10^{-25}\, \mathrm{m}^{-1} \nonumber\,, \\  
    &\overline{q}^z = ( +1.76 \pm 1.41 ) \times 10^{-25}\, \mathrm{m}^{-1}\,.
    \end{align}
Finally, for the 10-day interval from July 20th to 29th, 2023, we can get even tighter constraints:
\begin{align}
    &\overline{q}^x = (+ 6.18\pm 6.74)\times 10^{-26}\, \mathrm{m}^{-1}\nonumber\,, \\ 
    &\overline{q}^y = (+ 2.97 \pm 5.88) \times 10^{-26}\, \mathrm{m}^{-1} \nonumber\,, \\  
    &\overline{q}^z = (-5.04 \pm 6.59 ) \times 10^{-26}\, \mathrm{m}^{-1}\,,
\end{align}
where \(x, y, z\) refers to the directions of the coordinate axes of the ECI frame. 
Note that we have used ($\Delta f = 0.185 ~\text{mHz}$) when calculating the variance in Eq.~(\ref{variance}), which corresponds to the average over (40) adjacent data points. 
As expected, the errors decrease as we use longer time intervals (see Fig~\ref{fig2}), but there is no significant deviation from zero, which suggests that there is no systematic error at this level of precision. 

We can now translate this result into bounds on \(g_a \dot\phi\).
By using Eq.~(\ref{variance}), we have
\begin{align}
    |g_a \dot{\phi}| \lesssim 10^{-6} \, \mathrm{yr}^{-1}\,,
\end{align}
for a 1-day observation period and 
\begin{equation}
    |g_a \dot{\phi}| \lesssim 10^{-7} \, \mathrm{yr}^{-1}\,,
\end{equation}
for a 10-day observation period. 
 The bounds obtained here are worse than the naive bound in Eq.~(\ref{eq13}) imposed by atomic clocks on Earth. 
 However, as we noted before, the bound in Eq.~(\ref{eq13}) from terrestrial atomic clocks does not apply if the coupling to \(\phi(t)\) does not change the ratio of different atomic transition frequencies. 

 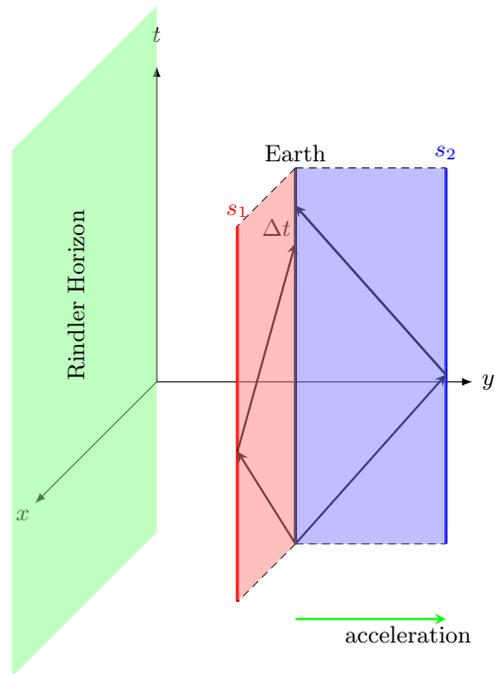
\begin{figure}
\centering
\begin{tikzpicture}[pics/3d axes/.style={code={%
 \draw[-latex] (0,0,0) -- (#1,0,0) node[pos=1.05]{$y$}; 
 \draw[-latex] (0,0,0) -- (0,#1,0) node[pos=1.1]{$t$}; 
 \draw[-latex] (0,0,0) -- (0,0,#1) node[pos=1.1]{$x$}; 
}}]

 \pic{3d axes=4.2};
 \draw[-, very thick](3, -1, 3)--(3, 4, 3);
\node (A) at (3, 4.2, 3) {Earth};

 \draw[-, very thick, red](3, -1, 5)--(3, 4, 5);
\node (B) at (3, 4.2, 5) {$\textcolor{red}{s_1}$};

 \draw[-, very thick, blue](5, -1, 3)--(5, 4, 3);
\node (C) at (5, 4.2, 3) {$\textcolor{blue}{s_2}$};

\draw[->, thick](3, -1, 3)--(3, 1, 5);

\draw[->, thick](3, 1, 5)--(3, 3, 3);

\draw[->, thick](3, -1, 3)--(5, 1.25, 3);

\draw[->, thick](5, 1.25, 3)--(3, 3.5, 3);

\draw[-, thick](5, 1.25, 3)--(3, 3.5, 3);

\draw[-, densely dashed, black](3, 4, 3)--(3, 4, 5);

\draw[-, densely dashed, black](3, -1, 3)--(3, -1, 5);

\draw[-, densely dashed, black](3, 4, 3)--(5, 4, 3);

\draw[-, densely dashed, black](3, -1, 3)--(5, -1, 3);

\node (D) at (2.75, 3.2, 3){$\Delta t$};

\fill [red!50,fill opacity=.5] (3, 4, 3) -- (3, 4, 5) -- (3, -1, 5) -- (3, -1, 3) -- cycle;

\fill [blue!50,fill opacity=.5] (3, 4, 3) -- (5, 4, 3) -- (5, -1, 3) -- (3, -1, 3) -- cycle;

\fill [green!50,fill opacity=.5] (0, -2, 0) -- (0, -2, 5) -- (0, 5, 5) -- (0, 5, 0) -- cycle;

\draw[semithick]
(-2, 0, -3) -- node[anchor=south, sloped] {Rindler Horizon}
cycle
;
\draw[->, thick, green](3, -2, 3)--(5, -2, 3);

\node (E) at (4.5, -2.2, 3){acceleration};

\end{tikzpicture}
    \caption{Universal coupling to a coherently evolving scalar field causes a uniform acceleration, $a$ along the direction of CMB dipole (Eq. \ref{eq:acceleration}). This can be measured by the precise timing of electromagnetic (null) signals that travel between Earth and satellites \(s_1\) (\(s_2\)), orthogonal to (along) the direction of acceleration. The plot shows these signals in the accelerated frame, which has a Rindler horizon, at a distance $1/a$ from Earth.  }
    \label{fig4}
\end{figure}

\begin{figure*}[t]
    \centering
    \includegraphics[scale=1.2]{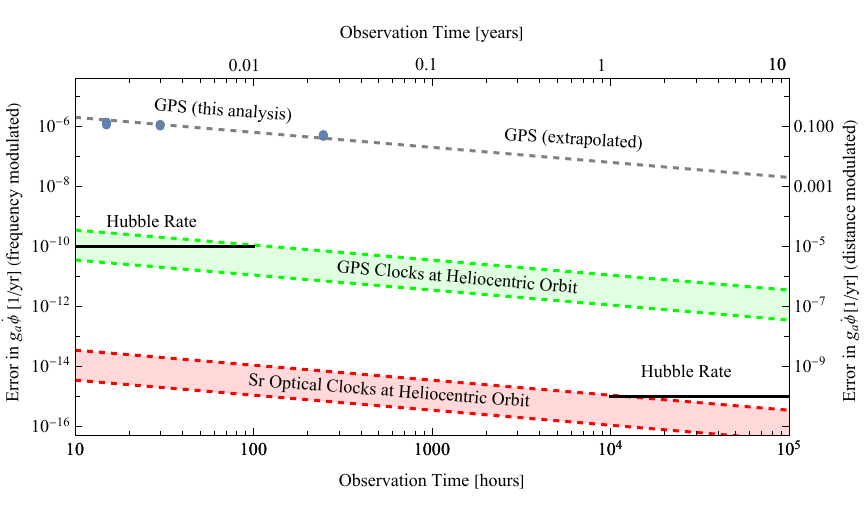}
    \caption{The error in \(g_a \dot{\phi}\) as a function of observation time. The left vertical axis depicts the scale in the ``dipole modulation of frequency" mechanism discussed in section~\ref{modulation to frequency}, while the right vertical axis depicts the scale in the ``dipole modulation of distance" mechanism discussed in section~\ref{modulation of distance}. The blue dots show the bounds obtained in this work, and the grey dotted line extrapolates these results for longer integration times. Better bounds can be obtained with data from GPS-type clocks (Sr optical clocks) in a heliocentric orbital pattern shown in green (in red). The horizontal black lines indicate the Hubble rate $\simeq 10^{-10}$ yr$^{-1}$ as possible cosmological targets for these measurements.} 
    \label{fig2}
\end{figure*}

\section{Dipole modulation of distance}\label{modulation of distance}

There is another effect, ignored in the previous calculation, that may cancel any observable frequency modulation of atomic clocks. If the same gradient that modulates the frequency of atomic clocks, modulates their mass, this will lead to an additional constant acceleration. As a result, in the rest frame of the clocks, the two gradient effects cancel. Nonetheless, photons that travel between Earth and GPS satellites do not experience this gradient, and thus their travel times will instead be modulated by the dipole, which we shall work out below.     

We shall then assume that, similar to the transition frequencies, the mass of the satellite will be modulated by the scalar field that changes over time. Under the influence of this field, the action of the satellite becomes
\begin{align}
    S=\int M_s (t) \sqrt{1-\frac{v^2}{c^2}} d t\,,
\end{align}
where \(v\) is the speed of the satellite relative to the scalar background. 

Similar to Eq.~(\ref{eq1}), we can write
\begin{align}
    \frac{\delta M_s}{M_s}=g_a \delta \phi\,.
\end{align}
To obtain the travel distance, the acceleration of the moving satellite is needed. By setting $\delta S=0$ we have
\begin{align}
    a^i=-\frac{\dot M_s}{M_s} { v}^i\,. \label{eq:acceleration}
\end{align}
If we assume that the coupling $\dot M_s/M_s $ with the scalar field is universal for all clocks (and satellites), it will be locally unobservable. However, given that the coupling does not affect the propagation of electromagnetic signals (as photons are massless), this acceleration can have an observable effect on timing for sending or receiving signals across well-separated clocks.   

To account for the constant acceleration of massive objects,  we can go the Rindler frame with the metric:
\begin{align}
    ds^2=-c^2(\kappa x-1)^2 dt^2+dx^2+dy^2,
\end{align}
Here, we can consider a 2D space since each satellite travels in its own orbital plane. Note that $\kappa=(\dot M_s/M_s)  v$ is the acceleration for the static observer at $x=0$.

By subtracting and integrating the time, we have
\begin{align}
    ct=\int_0^r \frac{dr'}{1- \kappa x}=r-\frac{1}{2}\kappa r^2 \cos(\theta)\,.
\end{align}
Here we have $x=r \cos\theta $, where $r$ is the radius of the satellite orbit, and $\theta$ is the relative angle between the satellite and the CMB dipole (assuming that it coincides with the direction of acceleration), as seen from Earth (see Fig.\ref{fig4}).
Therefore, the timing bias can be expressed as
\begin{align}
    d=-\frac{v_{\rm CMB}}{2 c^2}\frac{\dot M_s}{M_s}  \, r^2 \cos(\theta)\,.
\end{align}
In this equation, we have made the approximation \({v\approx v_{\rm CMB}}\). 
This is because we are considering the scalar background to be at rest in the cosmological frame, so we can approximate \(v\) as \(v_{\rm CMB}\), the speed of Earth in the cosmological frame, since \(v_{\rm sat}\) is much smaller in comparison. 

Taking a time derivative, we find:
\begin{align}
    \dot d=\frac{v_{\rm CMB} v_{\rm sat}}{2c^2}g_a\dot \phi\,  r\sin(\theta)\,. 
\end{align}
Comparing this to Eqs.~(\ref{eq6}-\ref{dnu}), we notice that
\begin{equation}
    \frac{\dot d_{\rm DM}}{\dot d_{\rm FM}} \sim \frac{v_{\rm sat}}{c}\,, 
\end{equation}
where the subscripts \(\rm DM\) and \(\rm FM\) stand for distance modulation and frequency modulation respectively.  This implies that the error for the distance modulated case is bigger by $\left(v_{\rm sat}/c\right)^{-1} = 10^5$ ($10^4$) for mid-Earth (Heliocentric) orbits, which is reflected on the right vertical axis of Figure~\ref{fig2}. 

\section{Conclusions and Outlook}
In this letter, we have analyzed the possibility of utilizing networks of quantum sensors to detect cosmological scalar fields predicted by various theoretical models. We have examined two mechanisms: the direct coupling of the scalar field to transition frequencies of atomic clocks and the modulation of distance resulting from a universal coupling between a scalar field and the effective mass of matter. We then used publicly available GPS timing bias and position data for primary data analysis. The results of these analyses have shown that, given the current GPS network and relatively short observation period, we cannot yet obtain interesting bounds on the existence of the cosmic scalar background. The error bound imposed by such data analysis seems to be worse than the naive upper bound implied by \cite{godun2014frequency}, although that upper bound is not applicable to a universal frequency modulation. 

There are a few ways to improve the bound of \(g_a \dot{\phi}\) imposed by orbital atomic clock networks. First, a longer observation period, such as 1-10 years, would improve this bound. This is in principle feasible at present, since GPS timing bias and position data are publicly available in \cite{NASAdata4}. However, we defer a full analysis of this data to future work, due to its large volume, and especially because some of the datasets have quality issues, such as missing data points. 
In addition to this, the precision of the atomic clocks currently used in GPS satellites is much worse than that of the newest terrestrial clocks \cite{s23135998}. Therefore, technological advances and the deployment of orbital atomic clock networks with better clocks in the future will continue to improve this bound. Lastly, the bound can be improved by deploying an orbital sensor network with a much larger orbital radius compared to GPS, such as a heliocentric orbital pattern. In Figure~\ref{fig2}, we show how the bound on \(g_a \dot{\phi}\) 
decreases for several network configurations. 

We should note that, while we have focused on the statistical error budget for detection of coupling to a cosmological scalar fields, potential systematic errors in modelling standard physics may further contaminate such endeavor, and thus any claim of detection (rather than upper limit) will require a careful accounting of such possible degeneracies.

A clear error target for a cosmological scalar background is \(g_a \dot{\phi}\sim 10^{-10}~\text{yr}^{-1}\), which implies an ${\cal O}(1)$ change in mass over a Hubble time. 
As can be seen in Figure \ref{fig2}, this is in principle realizable for both mechanisms, by deploying GPS-type atomic clocks and Sr lattice optical clocks at mid-Earth/heliocentric orbits and observing for a sufficiently long period. Although such detection networks might not be realistic in the near future, the improvement of current technologies, such as the deployment of the new generation of GPS satellites, will continue to improve the current result. Moreover, NASA has launched a series of missions aiming to send high-sensitivity atomic clocks into space, namely, the Deep Space Atomic Clock Missions \cite{NASAdeepspace}. Its first stage, which included a mercury ion clock on a low-Earth-orbit satellite, was terminated on September 18, 2021. A follow-up mission to Venus will be launched in 2028, which can lead to possible improvements of the upper bound on \(g_a \dot{\phi}\) if its timing-bias data can be accessed for data analysis. 

\begin{acknowledgements}
    We acknowledge the support of the Undergraduate Student Research Award from the Natural Sciences and Engineering Research Council of Canada (NSERC).  This research was funded thanks in part to the Canada First Research Excellence Fund through the Arthur B. McDonald Canadian Astroparticle Physics Research Institute, the Natural Sciences and Engineering Research Council of Canada, and the Perimeter Institute. Research at Perimeter Institute is supported in part by the Government of Canada through the Department of Innovation, Science and Economic Development and by the Province of Ontario through the Ministry of Colleges and Universities. 
\end{acknowledgements}

\bibliography{refs}
\end{document}